\documentclass[%
 reprint,
 amsmath,amssymb,
 aps,
 superscriptaddress,
 prd,
]{revtex4-2}

\usepackage[colorlinks,
           bookmarks=true,
           linkcolor=blue,
           urlcolor=blue, 
            anchorcolor=black,
            citecolor=blue
            ]{hyperref}

\usepackage{multirow}
\usepackage{graphicx}% Include figure files
\usepackage{dcolumn}% Align table columns on decimal point
\usepackage{bm}% bold math
\usepackage{float}
\usepackage{exscale}
\usepackage{relsize}

\begin{document}

%\preprint{APS/123-QED}

\title{Inclusive $J/\psi$ productions in pp collisions at $\sqrt{s}=$ 5.02, 7, and 13 TeV with the PACIAE model}% 
 
\author{Jin-Peng Zhang}
\affiliation{School of Physics and Information Technology, Shaanxi Normal University, Xi'an 710119, China}
\author{Guan-Yu Wang}
\affiliation{School of Physics and Information Technology, Shaanxi Normal University, Xi'an 710119, China}
\author{Wen-Chao Zhang}
\email{wenchao.zhang@snnu.edu.cn (corresponding author)}
\affiliation{School of Physics and Information Technology, Shaanxi Normal University, Xi'an 710119, China}

\author{Bo Feng}
\affiliation{School of Physics and Information Technology, Shaanxi Normal University, Xi'an 710119, China}
\author{An-Ke Lei}
\affiliation{School of Physics and Electronic Science, Guizhou Normal University, Guiyang, 550025, China}
\author{Zhi-Lei She}            
\affiliation{Wuhan Textile University, Wuhan 430200,
            China}
\author{Hua Zheng}
\affiliation{School of Physics and Information Technology, Shaanxi Normal University, Xi'an 710119, China}            
\author{Dai-Mei Zhou}
\email{zhoudm@mail.ccnu.edu.cn}
\affiliation{Key Laboratory of Quark and Lepton Physics (MOE) and Institute of
            Particle Physics, Central China Normal University, Wuhan 430079,
            China}            
\author{Yu-Liang Yan}
\affiliation{China Institute of Atomic Energy, P. O. Box 275 (10), Beijing
            102413, China}
\author{Ben-Hao Sa}
\email{sabhliuym35@qq.com} 
\affiliation{China Institute of Atomic Energy, P. O. Box 275 (10), Beijing
            102413, China}   
\affiliation{Key Laboratory of Quark and Lepton Physics (MOE) and Institute of
            Particle Physics, Central China Normal University, Wuhan 430079,
            China}
\date{\today}% It is always \today, today,
             %  but any date may be explicitly specified

\begin{abstract}

We investigate the inclusive $J/\psi$ production in proton-proton (pp) collisions at center-of-mass energies $\sqrt{s} = 5.02$, 7, and 13 TeV using the PACIAE 4.0  model. This model extends PYTHIA 8.3 by incorporating partonic and hadronic rescatterings before and after hadronization, respectively. Compared to our earlier study [K.-F. Ye et al., Phys. Rev. C 109, 035201 (2024)], which considered only the color-singlet processes, the present work includes both the color-singlet and color-octet contributions within the non-relativistic QCD (NRQCD) framework. In addition to NRQCD, we also consider the contributions from the cluster collapse and weak decays of $b$-hadrons. We find that the simulated inclusive $J/\psi$ transverse momentum differential cross sections agree  well with the experimental data at both the middle and forward rapidities. We provide a quantitative analysis of the relative contributions for different production mechanisms and their energy and rapidity dependence  in the inclusive $J/\psi$ production. Furthermore, we offer a  quantification of the relative contributions of the various components and their energy and rapidity dependence  in the NRQCD channels, including the direct production via the hard scattering and the feed-down from the decays of heavier charmonium states such as $\psi(2S)$, $\chi_{c0}$, $\chi_{c1}$, and $\chi_{c2}$. Finally, we examine the effects of partonic and hadronic rescatterings and offer the quantitative estimate of their impact on the $J/\psi$ production. These results are entirely new and represent a significant step forward in understanding the mechanisms of the inclusive $J/\psi$ production in high-energy pp collisions.

\end{abstract}

\maketitle

\section{\label{sec:intro}Introduction}

$J/\psi$, the lightest vector charmonium state composed of a charm and anti-charm quark pair, serves as a critical probe in heavy-ion physics. Its suppression in nucleus-nucleus collisions was originally proposed as a signature of the quark-gluon plasma \cite{probe_1}. However, this suppression can also be attributed to cold nuclear matter effects, such as the modification of nuclear parton distribution functions (PDFs) \cite{cme_1, cme_2}. In order to disentangle these hot and cold nuclear matter effects, a precise knowledge of the $J/\psi$ production in the absence of a nucleus in the initial state is indispensable. The $J/\psi$ yields in proton-proton (pp) collisions provide an essential reference for quantifying nuclear modifications in both proton-nucleus and nucleus-nucleus collisions.

In pp collisions at the LHC, the $J/\psi$ production predominantly originates from the hard scattering of two gluons, which generates a charm-anticharm ($c\bar{c}$) pair. A color-singlet $c\bar{c}$ pair may hadronize directly into a $J/\psi$, whereas a color-octet pair must first emit soft gluons to transition to the physical color-singlet state. While the initial gluon fusion process is calculable within perturbative QCD (pQCD), the subsequent hadronization remains a non-perturbative phenomenon that cannot be derived directly from  pQCD. Several theoretical approaches, such as the color-singlet model \cite{csm_model}, the non-relativistic QCD (NRQCD) \cite{NRQCD_1, NRQCD_2}, and the color evaporation model \cite{CEM_1,CEM_2}, have been proposed to describe  the $J/\psi$ production. A key distinction among these models lies on their description of the non-perturbative evolution of the initial $c\bar{c}$ pair into the observed $J/\psi$ bound state. The production of $J/\psi$ mesons has also been studied using Monte Carlo simulations \cite{mc_1, mc_2, mc_3, mc_4, Urqmd_1, sa_1, sa_2}.  According to Refs. \cite{mc_1, mc_2, mc_3, mc_4}, PYTHIA 8.2 \cite{pythia_8} successfully reproduces the positive correlation between the $J/\psi$ yield and the charged particle multiplicity ($dN_{\rm ch}/d\eta$) observed in pp collisions, whereas PYTHIA 6.4 \cite{pythia_6} predicts a negative correlation. This discrepancy may be attributed to differences in the treatment of multiparton interactions across the two PYTHIA versions. In Ref. \cite{mc_2}, the EPOS3 event generator \cite{epos_1, epos_2} was also employed to investigate the correlation between $J/\psi$ production and $dN_{\rm ch}/d\eta$. Meanwhile, Ref. \cite{Urqmd_1} applied a modified ultra-relativistic quantum molecular dynamics (UrQMD) transport model \cite{Urqmd_2, Urqmd_3} to examine $J/\psi$ suppression in high-multiplicity pp collisions at $\sqrt{s} = 7\ \text{TeV}$. In Refs. \cite{sa_1, sa_2}, the JPACIAE model was used to simulate the $J/\psi$ production in ultra-relativistic nucleus-nucleus collisions.

In our previous study \cite{Jpsi_paciae_22}, we used the parton and hadron cascade model PACIAE 2.2a \cite{paciae_22}, based on PYTHIA 6.4 \cite{pythia_6}, to investigate direct $J/\psi$ production in pp collisions at the LHC energies. In that work, only color-singlet processes such as $gg\rightarrow J/\psi g$, $gg\rightarrow \chi_{c0} g$, $gg\rightarrow \chi_{c1} g$, $gg\rightarrow \chi_{c2} g$, $gg\rightarrow \chi_{c1}$, $gg\rightarrow \chi_{c2}$, and $gg\rightarrow J/\psi \gamma$ were taken into account, while color-octet mechanisms were excluded. The higher-mass excited states $\chi_{c0}$, $\chi_{c1}$, and $\chi_{c2}$ were included via their subsequent decays into $J/\psi$. However, the feed-down from other excited states, such as $\psi(2S)$ and $\psi(3770)$, were not considered.

%since the $\psi(2S)$ production channel is not available in the default processes of PYTHIA 6.4, its feed-down contribution to $J/\psi$ was not considered.

As a complementary study to our previous work in Ref. \cite{Jpsi_paciae_22}, we investigate the inclusive $J/\psi$ production at both middle and forward rapidities in pp collisions at $\sqrt{s} = 5.02$, 7, and 13 TeV using the parton and hadron cascade model PACIAE 4.0 \cite{paciae_40}. Based on PYTHIA 8.3 \cite{pythia_83}, this model incorporates the partonic rescattering (PRS) before hadronization and the hadronic rescattering (HRS) afterward. The inclusive $J/\psi$ yield comprises both prompt and non-prompt contributions. The prompt component includes directly produced $J/\psi$ as well as feed-down from decays of heavier charmonium states such as $\psi(2S)$ and $\chi_c$, while the non-prompt component stems from the weak decay of $b$-hadrons \cite{Jpsi_5.02_mid}. In the model, the prompt $J/\psi$ production proceeds via the leading-order NRQCD channels including both the color-singlet and color-octet  contributions.  The prompt $J/\psi$ can also be generated through the cluster collapse mechanism \cite{cluster_collapse} during hadronization. This occurs when a charm quark and its  antiquark, being sufficiently close in phase space, form a quarkonium bound state directly, as the string potential energy is insufficient to produce a light quark–antiquark pair. The simulation transverse momentum ($p_{\rm T}$) spectra of the inclusive $J/\psi$ are compared with experimental data \cite{Jpsi_5.02_mid, Jpsi_7_mid, Jpsi_13_mid, Jpsi_5.02_13_fwd, Jpsi_7_fwd}. We also present the relative contributions of the NRQCD, the cluster collapse, and the non-prompt processes as well as their energy and rapidity dependence in the inclusive $J/\psi$ productions. Furthermore, in the NRQCD process, the fractions of the direct production from the parton hard scattering as well as the production from the decays of $\psi(2S)$, $\chi_{c0}$, $\chi_{c1}$, and $\chi_{c2}$ are quantified and their energy and rapidity dependence is provided. Finally, the effects of partonic and hadronic rescatterings on the $J/\psi$ production are systematically examined and quantitatively estimated.

The organization of this paper is as follows. Section \ref{sec:model} commences with a description of the PACIAE 4.0 model. This is followed   by a presentation of the results and a comprehensive discussion in section \ref{sec:results}. The paper concludes with a summary in section \ref{sec:conclusions}.

\section{\label{sec:model}The PACIAE model }
The PACIAE 4.0 model  \cite{paciae_40} extends PYTHIA 8.3 \cite{pythia_83} by incorporating partonic rescattering before hadronization and hadronic rescattering afterward. It describes ultra-relativistic pp collisions through four sequential stages: parton initialization, partonic rescattering, hadronization, and hadronic rescattering.

In the initial stage, the partonic state is generated by first disabling string fragmentation and then breaking diquarks and anti-diquarks into constituent partons. Along with initial- and final-state radiation, the system undergoes $2\rightarrow 2$ hard rescatterings, modeled using leading-order pQCD parton–parton cross sections \cite{cs_1, cs_2}. A $K$-factor is empirically applied as a multiplicative correction to the hard scattering cross sections \cite{paciae_30}. Following the partonic rescattering , the hadronization proceeds via either Lund string fragmentation \cite{pythia_6} or coalescence \cite{paciae_30}. In this work, the Lund string fragmentation scheme is employed. The subsequent hadronic rescattering  stage is implemented via a double loop over all hadron pairs $i$ and $j$. The minimum approach distance $D$ between their straight-line trajectories is computed, and a collision is considered possible if $D\leq \sqrt{\sigma^{\rm tot}_{ij}/\pi}$, where $\sigma^{\rm tot}_{ij}$ denotes the total cross section for the pair. The corresponding collision time $t_{ij}$ is then evaluated \cite{paciae_20}. An initial hadron–hadron collision time list is constructed for all possible pairs. The collision with the shortest time is selected and executed probabilistically, after which both the hadron list and collision time list are updated \cite{paciae_30}. This procedure repeats until the collision time list is empty, marking the onset of the kinetic freeze-out.

For $J/\psi$ hadronic rescattering, in addition to the elastic processes, the following inelastic channels are incorporated \cite{paciae_30, Jpsi_rescattering}:
\begin{eqnarray*}
J/\psi+n\rightarrow \Lambda_c^+ +D^-, \hspace{0.4cm}
J/\psi+n\rightarrow \Sigma_c^+ +D^-, \\[0.8mm]
J/\psi+n\rightarrow \Sigma_c^0 +\bar{D}^0, \hspace{0.4cm}
J/\psi+p\rightarrow \Lambda_c^+ +\bar{D}^0,\\[0.8mm]
J/\psi+p\rightarrow \Sigma_c^+ +\bar{D}^0, \hspace{0.4cm}
J/\psi+p\rightarrow \Sigma_c^{++} +D^-,\\[0.8mm]
J/\psi+\pi^+\rightarrow D^+ +\bar{D}^{*0}, \hspace{0.4cm}
J/\psi+\pi^-\rightarrow D^0 +D^{*-},\\[0.8mm]
J/\psi+\pi^0\rightarrow D^0 +\bar{D}^{*0}, \hspace{0.4cm}
J/\psi+\pi^0\rightarrow D^+ +D^{*-},\\[0.8mm]
J/\psi+\rho^{+}\rightarrow D^+ +\bar{D}^{0}, \hspace{0.4cm}
J/\psi+\rho^-\rightarrow D^0 +D^{-},\\[0.8mm]
J/\psi+\rho^0\rightarrow D^0 +\bar{D}^{0}, \hspace{0.4cm}
J/\psi+\rho^0\rightarrow D^+ +D^{-}.
\end{eqnarray*}
Other approaches, such as the comover interaction model, also incorporate the hadronic rescattering effects \cite{comover_1,comover_2,comover_3, comover_4, comover_5, comover_51, comover_6, comover_8}.

\begin{figure*}[htbp]
\includegraphics[scale=0.46]{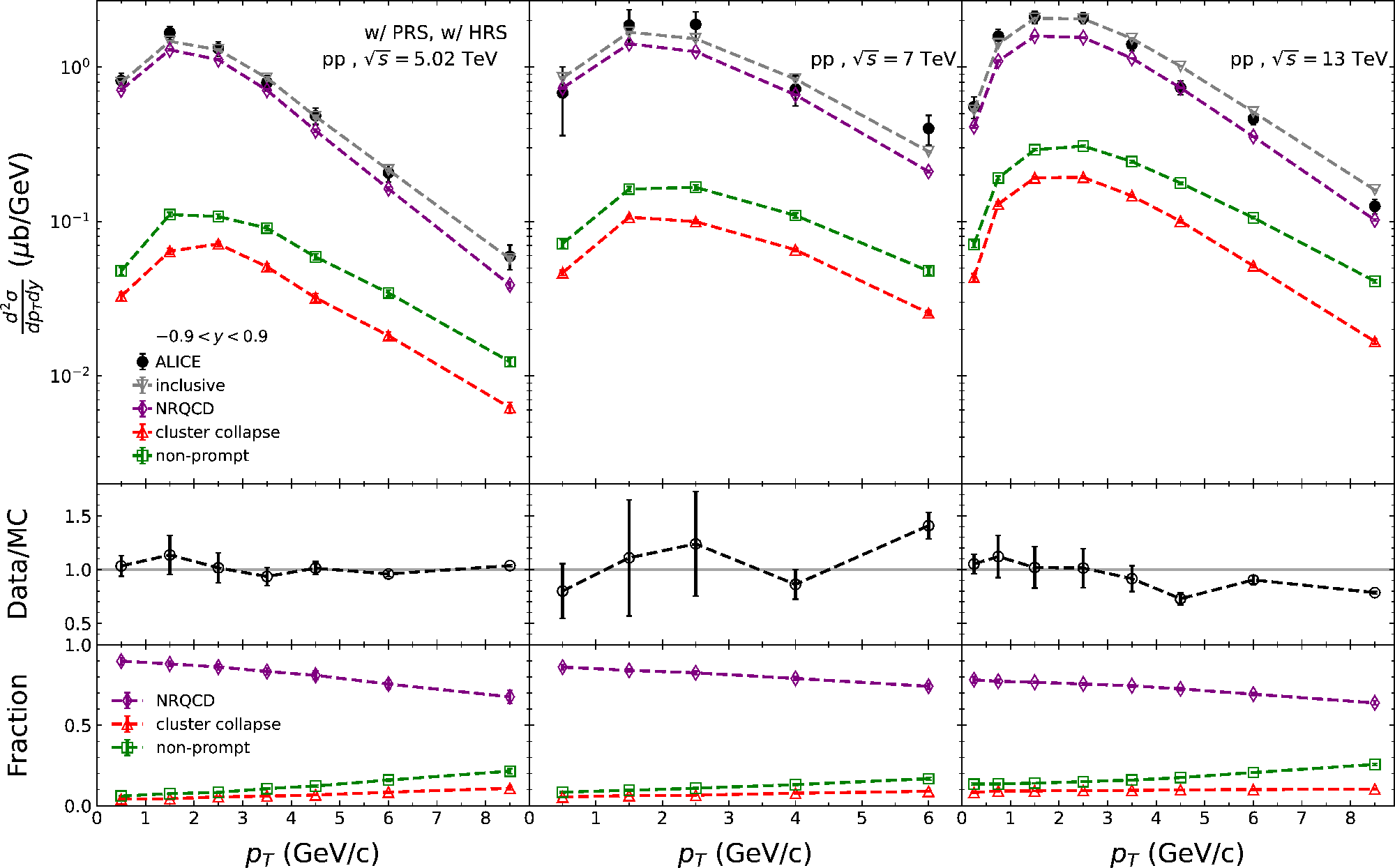}% Here is how to import EPS art
\caption{\label{fig:all_ph_mid} Upper row panels: the $p_{\rm T}$-differential cross sections of $J/\psi$ from the inclusive process (downward triangles), the NRQCD (diamonds), the cluster collapse (upward triangles), and the non-prompt process (squares)  at mid-rapidity in pp collisions at $\sqrt{s}=$ 5.02, 7, and 13 TeV in the PACIAE model. The solid circles are experimental data taken from Refs. \cite{Jpsi_5.02_mid, Jpsi_7_mid, Jpsi_13_mid}. Middle row panels: the ratio between the experimental data and the inclusive $J/\psi$ results from the PACIAE model. Lower row panels: the fractions for the NRQCD, cluster collapse, and non-prompt components in the inclusive $J/\psi$ productions.  }
\end{figure*}

\begin{figure*}[htbp]
\includegraphics[scale=0.46]{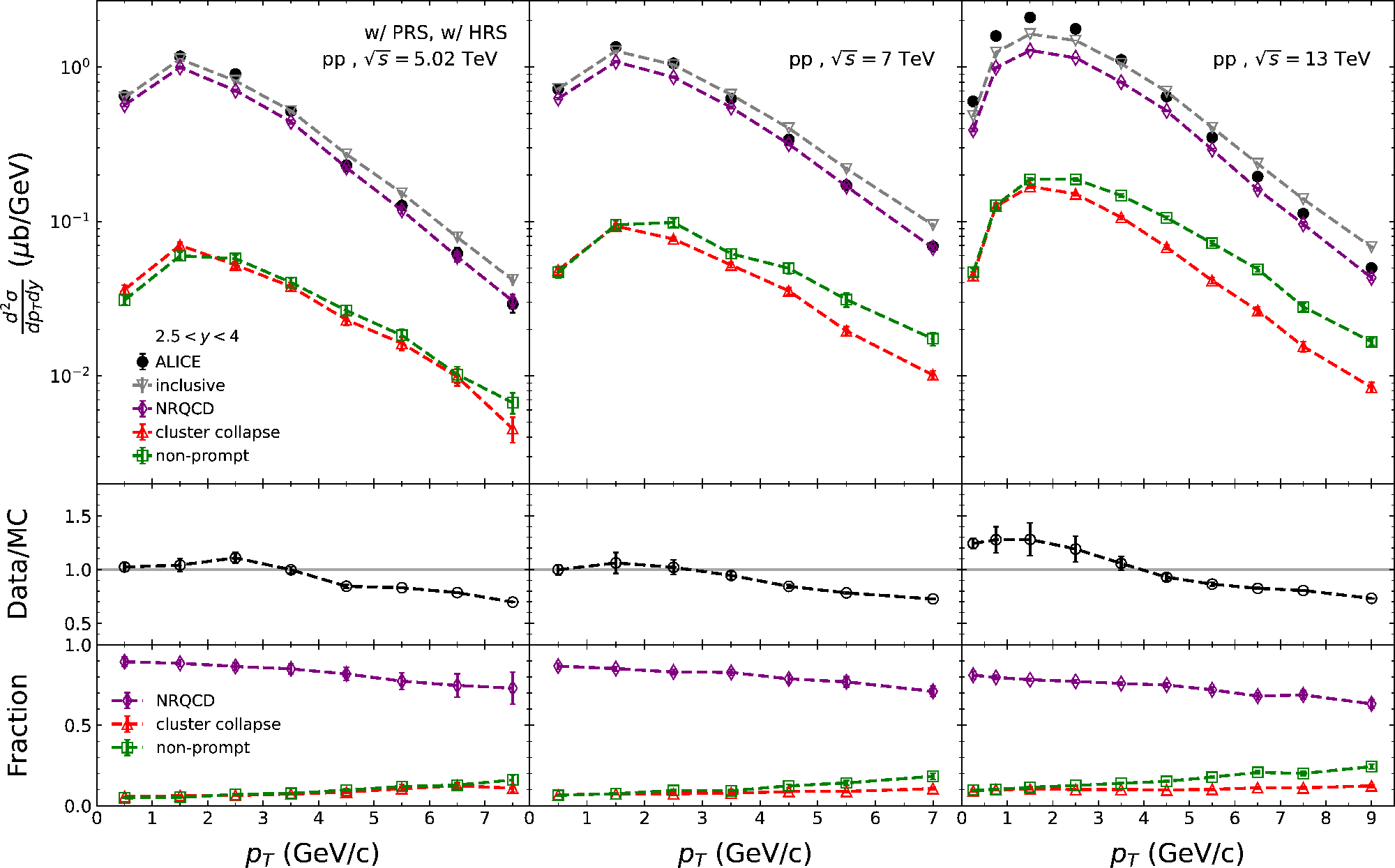}% Here is how to import EPS art
\caption{\label{fig:all_ph_for} Similar  as that in Fig. \ref{fig:all_ph_mid}, but for the $p_{\rm T}$-differential inclusive  $J/\psi$  cross sections at forward rapidity in pp collisions at $\sqrt{s}=$ 5.02, 7, and 13 TeV.  The experimental data are taken from Refs.~\cite{Jpsi_5.02_13_fwd, Jpsi_7_fwd}.  }
\end{figure*}

%This is followed by the hadronic rescattering stage, where hadrons undergo further two-body collisions \cite{two_body} until kinetic freeze-out is reached.

\section{\label{sec:results}Results and discussions}
We generated 400 million inelastic non-diffractive events in pp collisions at $\sqrt{s} = 5.02$, 7, and 13 TeV with the PACIAE 4.0 model. The model employed the default parameter set from the PYTHIA 8.3 Monash tune \cite{Monash_tune}, with the exception of the $K$ factor. To align the simulation results with experimental data, the $K$ factors were adjusted to 0.68, 0.71, and 0.75 for $\sqrt{s} = 5.02$, 7, and 13 TeV, respectively. The simulation includes critical mechanisms such as multiple-parton interactions (MPIs) \cite{MPI} and color reconnections (CR) \cite{CR_1}. MPIs comprise multiple independent $2\rightarrow 2$ partonic scatterings that primarily contribute to the underlying event. In high-energy pp collisions, however, a substantial fraction of MPIs can be sufficiently hard to produce a $J/\psi$, thereby increasing the total $J/\psi$ yield in events with high charged-particle multiplicity. These MPIs lead to the formation of overlapping color strings. The CR mechanism reduces the overall string length, effectively leading to a reduction of the total multiplicity. It is found that the MPI with the CR well reproduces the dependence of the mean transverse momentum on the charged particle multiplicity in pp collisions observed by the ALICE collaboration at the LHC \cite{CR_2}.

The upper row panels of Fig. \ref{fig:all_ph_mid} present the $p_{\rm T}$-differential inclusive $J/\psi$ cross sections at mid-rapidity in pp collisions at $\sqrt{s} = 5.02$, 7, and 13 TeV. Solid circles denote the experimental data, while open downward triangles represent the inclusive $J/\psi$ results from the PACIAE model, which incorporates PRS and HRS effects. In the PACIAE model, the $J/\psi$ cross section is obtained by normalizing the per-event differential $J/\psi$ yield to the nucleon–nucleon inelastic non-diffractive cross section ($\sigma_{\rm ND}^{\rm NN}$) \cite{Jpsi_7_mid}, as given by
\begin{eqnarray}
\frac{d^2\sigma_{J/\psi}}{dp_{\rm T}dy}=\frac{\sigma_{\rm ND}^{\rm NN}}{N_{\rm ND}^{\rm NN}}\frac{d^2N_{J/\psi}}{dp_{\rm T}dy},
\label{eq:yield_xs_transform}
\end{eqnarray}
where $N_{\rm ND}^{\rm NN}$ is the total number of inelastic non-diffractive events. The values of $\sigma_{\rm ND}^{\rm NN}$ for pp collisions at $\sqrt{s} = 5.02$, 7, and 13 TeV are 47.72 mb, 50.76 mb, and 56.42 mb, respectively, as calculated by PACIAE 4.0. These values are consistent with the estimations from the experimental measurements \cite{MB_cs}. The middle row panels display the ratios between the data and the inclusive results from the model, showing that the PACIAE model describes the data reasonably well. Except for the last data point at $\sqrt{s} = 7$ TeV, the discrepancy between data and model is within 30$\%$. The upper row panels also show the $p_{\rm T}$-differential cross sections for the $J/\psi$ mesons from the NRQCD channels (empty diamonds), the cluster collapse (empty upward triangles), and the non-prompt process (empty squares). The lower row panels present their corresponding fractions in the inclusive $J/\psi$ production. The results indicate that the dominant contribution comes from the NRQCD production, while the fractions of $J/\psi$ from the cluster collapse and the non-prompt process are relatively small. The $p_{\rm T}$-average fractions of prompt and non-prompt components in inclusive $J/\psi$ production  at mid-rapidity in pp collisions at $\sqrt{s} = 5.02$, 7, and 13 TeV are summarized in Table \ref{tab:fractions_Jpsi_in_inclusive}. With increasing collision energy, the relative contributions of the cluster collapse and non-prompt processes to the inclusive $J/\psi$ production exhibit a growing trend, while the fraction originating from the NRQCD channel decreases. This behavior stems from the distinct energy dependence of the production cross sections in each channel. At LHC energies, the $J/\psi$ cross section from the NRQCD process grows relatively slowly, scaling roughly as $\sigma \propto s^{-\alpha}$ with $\alpha > 0$ \cite{NRQCD_1}. In contrast, the cross section for $b\bar{b}$ production rises more rapidly with energy than that for $c\bar{c}$ \cite{bbbar_ccbar_cs}. Consequently, although the absolute yield from NRQCD still increases, its relative contribution is gradually diluted by the faster-growing cluster collapse and non-prompt channels.
% Please add the following required packages to your document preamble:
% \usepackage{multirow}
%To facilitate comparison with the experimental data, the PACIAE model results are scaled by factors of 0.68, 0.74, and 0.78 for $\sqrt{s} = 5.02$, 7, and 13 TeV, respectively.

\begin{table}[]
\caption{The $p_{\rm T}$-average fractions for the  NRQCD, cluster collapse, and non-prompt components in the inclusive $J/\psi$ productions at both the middle (mid) and  forward (fwd)  rapidities in pp collisions at $\sqrt{s} = 5.02$, 7, and 13 TeV. The uncertainties quoted are statistical errors.  }\label{tab:fractions_Jpsi_in_inclusive}
\begin{ruledtabular}
\begin{tabular}{ccccc}
             $\sqrt{s}$      & \multirow{2}{*}{$y$}        & \multirow{2}{*}{NRQCD}  & \multirow{2}{*}{cluster collapse}  & \multirow{2}{*}{non-prompt}  \\
             (TeV)      &        &  &  &  \\\colrule
\multirow{2}{*}{5.02} & mid     &   $(84.92\pm0.87)\%$ &$(5.58\pm 0.12)\%$&   $(9.50\pm 0.17)\%$         \\
                      & fwd &   $(86.27\pm1.19)\%$ &$(6.86\pm 0.18)\%$&   $(6.87\pm 0.18)\%$           \\\colrule
\multirow{2}{*}{7}    & mid     & $(81.75\pm 0.58)\%$  &$(6.90\pm0.09) \%$&   $(11.34\pm 0.23)\%$         \\
                      & fwd & $(83.03\pm 0.75)\%$  &$(7.69\pm 0.13)\%$&   $(9.29\pm 0.27)\%$\\\colrule
\multirow{2}{*}{13}   & mid     & $(74.22\pm0.33) \%$  &$(9.53\pm 0.09) \%$&   $(16.26\pm 0.13)\%$         \\
                      & fwd & $(76.31\pm 0.44)\%$  &$10.20\pm 0.13)\%$ &   $(13.49\pm 0.15)\%$        
\end{tabular}
\end{ruledtabular}
\end{table}

The upper row panels of Fig. \ref{fig:all_ph_for} present the $p_{\rm T}$-differential inclusive $J/\psi$ cross sections at forward rapidity in pp collisions at $\sqrt{s} = 5.02$, 7, and 13 TeV. The markers are the same as those at mid-rapidity in Fig. \ref{fig:all_ph_mid}. The middle row panels show the ratio between the experimental data and the inclusive $J/\psi$ results from the PACIAE model.  These ratios indicate that the model agrees with the data within 30$\%$. We emphasize that the $J/\psi$ productions at middle and forward rapidities are studied simultaneously. In the high $p_{\rm T}$ region, the model overestimates the data for all the three energies.   The lower row panels show the production fractions of the NRQCD, the cluster collapse, and the non-prompt components in the inclusive $J/\psi$ yield, with their $p_{\rm T}$-average values at forward rapidity summarized in Table \ref{tab:fractions_Jpsi_in_inclusive}. At forward rapidity, the energy dependence of these fractions is consistent with that observed at mid-rapidity. However, their absolute values differ significantly: the contributions from both the NRQCD and cluster collapse processes are enhanced, while the non-prompt component is suppressed. This could be understood as follows. The forward rapidity region amplifies an asymmetric collision environment where the extremely high density of small-$x$ partons significantly enhances the $J/\psi$ productioin from the NRQCD and the  cluster collapse processes, while the scarcity of large-$x$ partons suppresses the bottom-quark production, which requires a higher energy threshold. This in turn reduces the non-prompt $J/\psi$ yield from the $b$-hadron weak decays.
%At forward rapidity, a very small-$x$ gluon from one proton collides with a relatively large-$x$ gluon from the other proton. 

%The gluon density at very small $x$ is extremely high.  This high-density, small-$x$ gluon participates directly in the hard scattering, significantly boosting the probability of the gluon-fusion process, thereby enhancing the $J/\psi$ production from the NRQCD process.
%Similar to NRQCD,  the $J/\psi$ production from the cluster collapse process also benefits from the high small-$x$ gluon density. However, the number of large-$x$ gluons capable of meeting the bbˉbbˉ production threshold is relatively scarce. This leads to a relative suppression of the bbˉbbˉ production cross section at forward rapidity, which in turn reduces the non-prompt J/ψJ/ψ yield from their decays.

%At forward rapidity, we primarily probe collisions between a small-$x$ parton and a large-$x$ parton, with $x=p_{\rm T}/\sqrt{s}e^{\pm y}$.

\begin{figure*}[]
\includegraphics[scale=0.53]{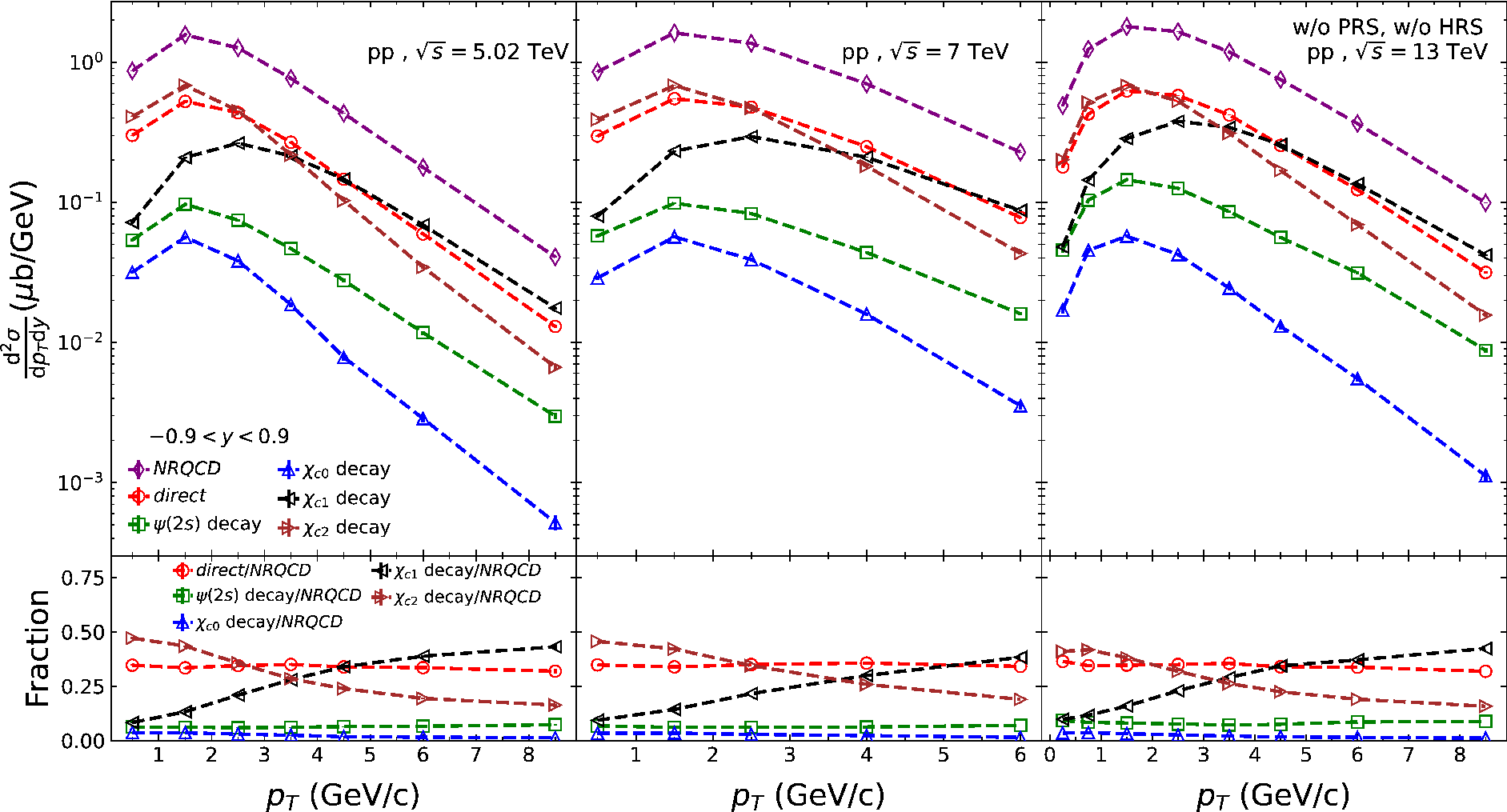}% Here is how to import EPS art
\caption{\label{fig:NRQCD_13_TeV_mid} Upper  panels: the $p_{\rm T}$-differential cross sections for $J/\psi$ from the direct production (circles) as well as the feed-down contributions  from the decays of $\psi(2S)$ (squares), $\chi_{c0}$ (upward triangles), $\chi_{c1}$ (left triangles), and $\chi_{c2}$ (right triangles) at middle rapidity in pp collisions at $\sqrt{s}=$ 5.02, 7, and 13 TeV. The $p_{\rm T}$-differential cross sections for $J/\psi$ from the total NRQCD processes is represented by the empty diamonds. Lower panels: the fractions of different components in the NRQCD processes.}
\end{figure*}

\begin{figure*}[]
\includegraphics[scale=0.53]{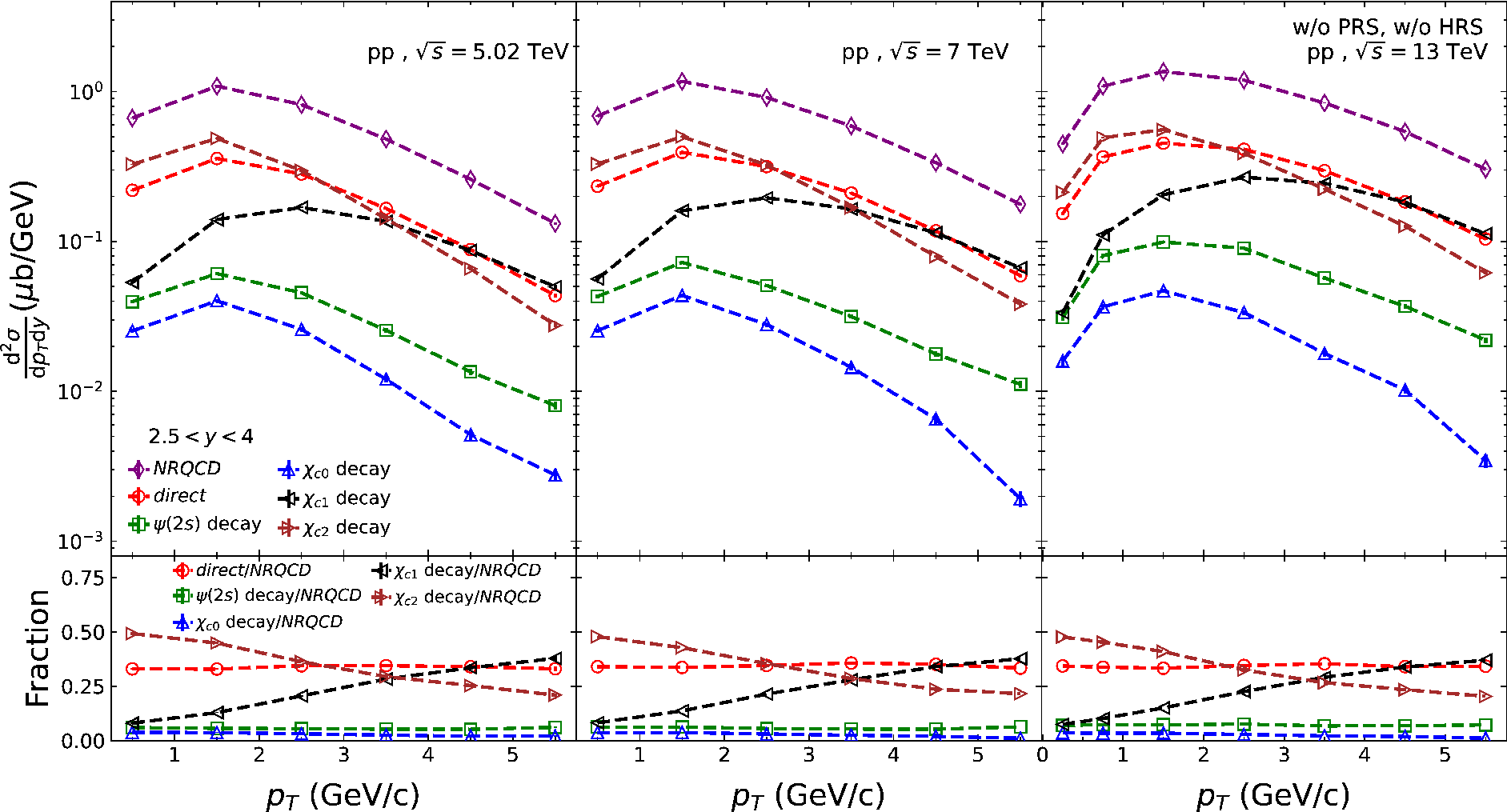}% Here is how to import EPS art
\caption{\label{fig:NRQCD_13_TeV_fwd} Similar  as that in Fig. \ref{fig:NRQCD_13_TeV_mid}, but for the $p_{\rm T}$-differential  cross sections for $J/\psi$ from different NRQCD processes at forward rapidity in pp collisions at $\sqrt{s}=$ 5.02, 7, and 13 TeV.  }
\end{figure*}

\begin{table*}[]
\caption{The $p_{\rm T}$-average fractions of $J/\psi$ from the direct production, and from the decays of $\psi(2S)$, $\chi_{c0}$, $\chi_{c1}$, and $\chi_{c2}$ in the NRQCD component at both the middle and  forward  rapidities in pp collisions at $\sqrt{s} =$ 5.02, 7, and 13 TeV. The uncertainties quoted are  statistical errors.} \label{tab:fractions_Jpsi_in_NRQCD}
\begin{ruledtabular}
\begin{tabular}{ccccccc}
   $\sqrt{s}$ (TeV)    &$y$      &direct         &  $\psi(2S)$       & $\chi_{c0}$ & $\chi_{c1}$ & $\chi_{c2}$ \\\colrule
\multirow{2}{*}{5.02}&mid     &   $(34.13\pm 0.14)\%$ &$(6.18\pm 0.05)\%$&   $(2.98\pm 0.04)\%$&     $(20.40\pm 0.10)\%$   &  $(36.30\pm 0.15)\%$ \\
                    &fwd     &   $(33.60\pm0.19) \%$ &$(5.61\pm 0.07)\%$&   $(3.23\pm 0.05)\%$ & $(18.44\pm 0.13)\%$  &   $(39.12\pm 0.21)\%$   \\\colrule
\multirow{2}{*}{7}   &mid     &  $(34.76\pm 0.14)\%$ &$(6.31\pm 0.05)\%$&   $(2.87\pm0.04)\%$&   $(21.09\pm 0.11)\%$   & $( 34.97\pm 0.14)\%$ \\
                    &fwd    &   $(34.35\pm0.19)\%$ &$(5.85\pm 0.07)\%$&   $(3.10\pm0.05)\%$ &  $(19.56\pm0.14)\%$  &   $(37.14\pm0.20)\%$  
\\\colrule
\multirow{2}{*}{13}  &mid    &   $(34.60\pm0.14)\%$ &$(8.01\pm0.06)\%$&   $(2.50\pm 0.03)\%$&   $(24.39\pm 0.11)\%$   &  $(30.51\pm0.13)\%$ \\
                     &fwd    &   $(34.10\pm0.17)\%$ &$(7.24\pm0.07)\%$&   $(2.70\pm0.04)\%$ &  $(22.61\pm0.13)\%$  &   $(33.35\pm0.17)\%$ 
\end{tabular}
\end{ruledtabular}
\end{table*}

As the yield of $J/\psi$ from  NRQCD includes the direct production from the parton hard scattering as well as the feed-down from decays of heavier charmonium states such as $\psi(2S)$, $\chi_{c0}$, $\chi_{c1}$, and $\chi_{c2}$, we investigate the relative contribution of each production mechanism. In the upper  panels of Figs. \ref{fig:NRQCD_13_TeV_mid} and \ref{fig:NRQCD_13_TeV_fwd}, we present the $p_{\rm T}$-differential inclusive $J/\psi$ cross sections from different NRQCD production mechanisms at both the middle and forward rapidities in pp collisions at  $\sqrt{s} =$ 5.02, 7, and 13 TeV from the PACIAE model. In this model, the PRS and HRS are deactivated because, when enabled, they technically prevent the tracing of the $J/\psi$'s mother particle. The empty circles, squares, upward triangles, left triangles, and right triangles represent the  $p_{\rm T}$ spectra of the $J/\psi$ mesons from the direct production, as well as the decays of $\psi(2S)$, $\chi_{c0}$, $\chi_{c1}$, and $\chi_{c2}$, respectively. The feed-down contributions from the decay of $h_c$ and $\psi(3770)$ are negligible, thus they are not presented in the figure. The total NRQCD yield, being the sum of all these components, is represented by empty diamonds. The corresponding production fractions are detailed in the lower panels. It is observed that at low  (high) $p_{\rm T}$ the dominant contributions come from the direct production and the $\chi_{c2}$ ($\chi_{c1}$) decay. The feed-down fractions from $\psi(2S)$ and $\chi_{c0}$ are relatively small. The differing contributions of various charmonium states to the inclusive $J/\psi$ yield stem primarily from the interplay between their distinct long-distance matrix elements (LDMEs) and the short-distance production mechanisms (color-singlet vs. color-octet). For the directly produced $J/\psi$,  at low $p_{\rm T}$ the color-singlet process, such as $gg\rightarrow J/\psi g$, is the largest contributor, while at high $p_{\rm T}$ the color-octet process becomes significant \cite{singlet_octet}. For the $\chi_{cJ}$ ($J=0,1,2$) states, the key to their relative sizes lies in the values of their color-octet LDMEs. The hierarchy of the LDMEs is $\chi_{c1}>\chi_{c2}>\chi_{c0}$ \cite{chi_c_LDMEs}. At low $p_{\rm T}$, the $\chi_{cJ}$ production is dominated by the gluon-gluon fusion, $gg\rightarrow \chi_{cJ}$. The spin structure of  $\chi_{c2}$ matches the total angular momentum of two gluons more favorably. This results in larger short-distance coefficients for the $\chi_{c2}$ production compared to $\chi_{c1}$. Even though $\chi_{c1}$'s LDME is intrinsically larger, the advantage of larger short-distance coefficients causes the absolute production cross section of $\chi_{c2}$ to surpass that of $\chi_{c1}$. At high $p_{\rm T}$, the production of $\chi_{cJ}$ via high-energy gluon fragmentation channel becomes important. It is easier for a high-energy gluon (spin-1) to transfer and reconfigure its angular momentum into $\chi_{c1}$ (spin-1) than into $\chi_{c2}$ (spin-2) during the fragmentation process. Thus, $\chi_{c1}$, with its larger color-octet long-distance matrix element and more favorable fragmentation function, surpasses $\chi_{c2}$ at high $p_{\rm T}$. For $\psi(2S)$, as a radial excitation, its wave function has a node at the origin, resulting in a much smaller cross section compared to the ground-state $J/\psi$ \cite{psi_2s}. For $\chi_{c0}$, the extremely small LDME strongly suppresses its color-octet production. Moreover, in the color-singlet channel, the leading-order process $gg\rightarrow \chi_{c0}$ is forbidden at tree level due to the angular momentum conservation. The $p_{\rm T}$-average fractions for all components in the NRQCD process at different energies are summarized in Table \ref{tab:fractions_Jpsi_in_NRQCD}. It is observed that the fraction of directly produced $J/\psi$ increases modestly with collision energy and appears to saturate at $\sqrt{s} =$ 13 TeV. Furthermore, the feed-down contributions from the $\psi(2S)$ and $\chi_{c1}$ decays grow with energy, whereas those from $\chi_{c0}$ and $\chi_{c2}$ exhibit a declining trend.  The observe energy dependence could be understood as follows. With increasing the collision energy, the dominant production mechanism evolves from the color-singlet hard scattering toward the color-octet fragmentation.  The vector nature and large associated LDME of $\chi_{c1}$ make it a primary beneficiary. Similarly, the favorable spin alignment of $\psi(2S)$ also allows it to gain from this evolution. In contrast, for $\chi_{c2}$, its spin-2 structure places it at an angular momentum disadvantage in fragmentation compared to $\chi_{c1}$. For $\chi_{c0}$, as its color-octet LDME is extremely small, its production relies mainly on the  color-singlet mechanism, causing its relative share to decline. These results provide strong observational grounds for using energy scans to separate and constrain the various LDMEs in NRQCD, particularly for distinguishing among the matrix elements for different $\chi_{cJ}$ states.  A clear rapidity dependence is also observed: at forward rapidity, the fractions originating from direct production as well as from $\psi(2S)$ and $\chi_{c1}$  decays are slightly suppressed relative to their values at mid-rapidity, while the contributions from $\chi_{c0}$ and $\chi_{c2}$ decays are moderately enhanced. This pattern is a direct consequence of the differing sensitivities of various production mechanisms to the parton momentum fraction $x$. The forward rapidity region suppresses the color-octet fragmentation that requires efficient production of a parent parton from the large-$x$ region, while  enhances the color-singlet hard scattering that can effectively utilize the extremely high small-$x$ gluon density.

\begin{table*}[]
\caption{The $p_{\rm T}$-average ratios of the $J/\psi$ cross section from the inclusive (NRQCD, cluster collapse, non-prompt) process with HRS-only, PRS-only, and both PRS and HRS to that with neither PRS nor HRS at both the middle and  forward rapidities in pp collisions at $\sqrt{s} =$ 13 TeV. The uncertainties quoted are  statistical errors.} \label{tab:PRS_HRS_ratio}
\begin{ruledtabular}
\begin{tabular}{cccccc}
                     &$y$      &inclusive         &  NRQCD       & cluster collapse & non-prompt  \\\colrule
\multirow{2}{*}{w/ PRS, w/o HRS}&mid     &   $(99.84\pm 0.40)\%$ &$(99.61\pm 0.46)\%$&   $(97.12\pm 1.25)\%$&     $(102.86\pm 1.06)\%$    \\
                    &fwd     &   $(99.11\pm0.51) \%$ &$(99.17\pm 0.58)\%$&   $(99.02\pm 1.56)\%$ & $(98.80\pm 1.43)\%$     \\\colrule
\multirow{2}{*}{w/ HRS, w/o PRS}   &mid     &  $(91.46\pm 0.37)\%$ &$(89.57\pm 0.42)\%$&   $(91.79\pm1.20)\%$&   $(101.01\pm 1.05)\%$    \\
                    &fwd    &   $(92.43\pm0.48)\%$ &$(91.54\pm 0.55)\%$&   $(91.14\pm1.47)\%$ &  $(98.96\pm1.43)\%$   
\\\colrule
\multirow{2}{*}{w/ PRS, w/ HRS}  &mid    &   $(92.27\pm0.37)\%$ &$(90.63\pm0.43)\%$&   $(90.31\pm 1.18)\%$&   $(102.02\pm 1.05)\%$    \\
                     &fwd    &   $(92.46\pm0.48)\%$ &$(91.77\pm0.55)\%$&   $(89.61\pm1.45)\%$ &  $(99.11\pm1.43)\%$  
\end{tabular}
\end{ruledtabular}
\end{table*}

%The feed-down fraction from $\psi(2S)$ is 6-8$\%$ at mid-rapidity and 8$\%$ at forward rapidity, while that from $\chi_{c0}$ is about 2$\%$ and 3$\%$, respectively. 
%is 6-8$\%$ at mid-rapidity and 5-7$\%$ at forward rapidity, while that from $\chi_{c0}$ is about 2$\%$ and 3$\%$, respectively.
%Similar results are also observed for the other two collision energies.

\begin{figure*}[htbp]
\includegraphics[scale=0.45]{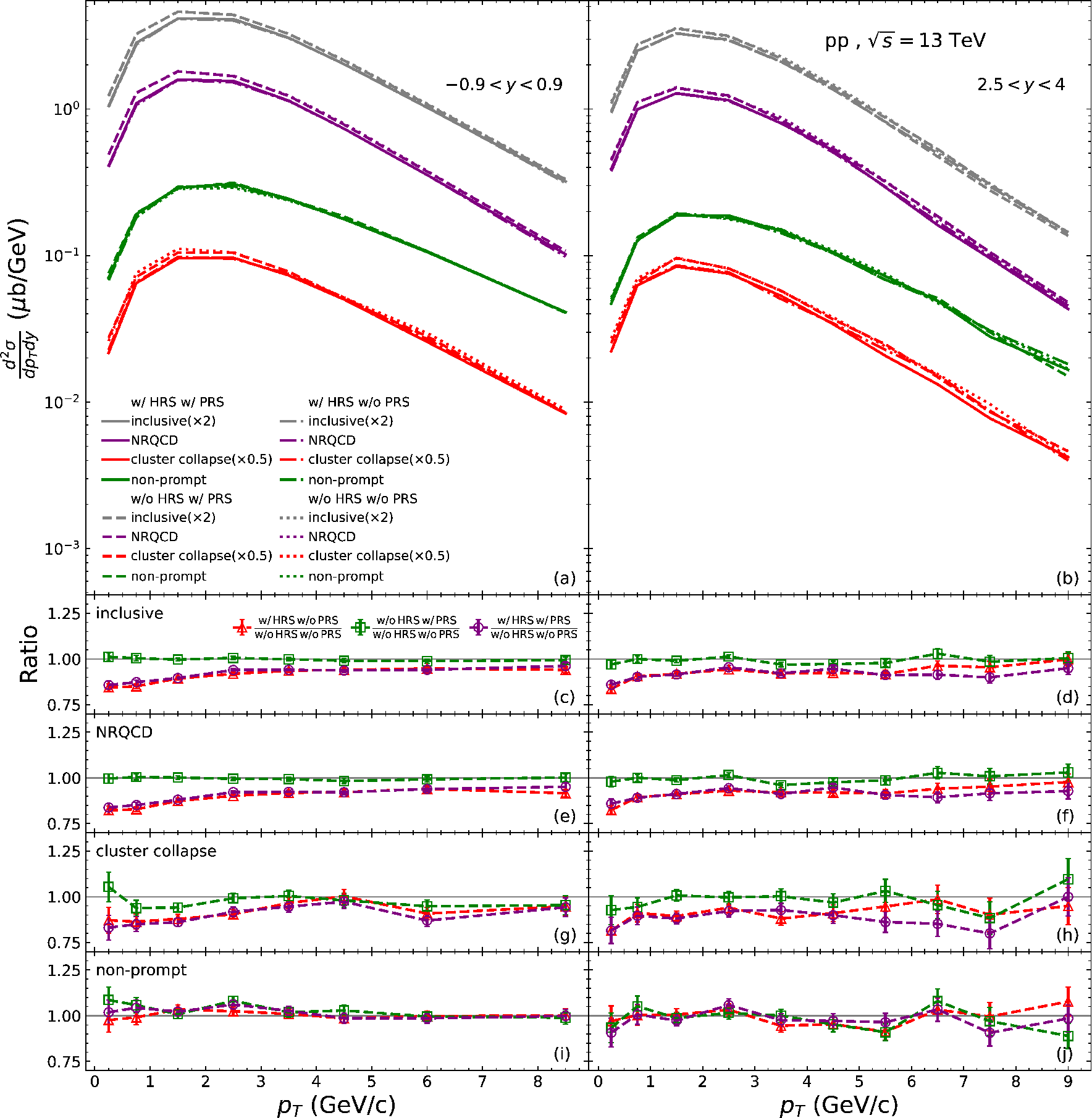}% Here is how to import EPS art
\caption{\label{fig:HRS_PRS} Panels (a) and (b): the $p_{\rm T}$-differential   cross sections of $J/\psi$ from from the inclusive process (grey curves)  as well as the NRQCD (red curves) , cluster collapse (green curves) and non-prompt (purple curves) processes at middle and forward rapidities in pp collisions at  $\sqrt{s} =$ 13 TeV in the PACIAE model under different rescattering scenarios: both PRS and HRS included (solid curves), HRS only  (dash-dotted curves), PRS only  (dashed curves), and neither PRS nor HRS, the  baseline (dotted curves). Panels (c) and (d) ((e) and (f), (g) and (h), (i) and (j)):  the ratios of the spectra from the inclusive (NRQCD, cluster collapse, non-prompt) process with HRS-only (upward triangles), PRS-only (squares), and both PRS and HRS (circels) to this baseline. }
\end{figure*}

Finally, we  investigate the effects of PRS and HRS on the $J/\psi$ production. As an example, in the panels (a) and (b) of Fig.  \ref{fig:HRS_PRS}, we present the $p_{\rm T}$-differential  $J/\psi$ cross sections at middle and forward rapidities in pp collisions at  $\sqrt{s} =$ 13 TeV in the PACIAE model under different rescattering scenarios. The grey, red, green, and purple curves represent the $p_{\rm T}$ spectra of  $J/\psi$ from the inclusive process  as well as the NRQCD, cluster collapse and non-prompt processes, respectively. The line styles denote the following rescattering configurations: solid curves (both PRS and HRS included), dash-dotted curves (HRS only), dashed curves (PRS only), and dotted curves (neither PRS nor HRS, the  baseline). Panels (c) and (d) ((e) and (f), (g) and (h), (i) and (j)) present the ratios of the spectra from the inclusive (NRQCD, cluster collapse, non-prompt) with HRS-only (empty upward triangles), PRS-only (empty squares), and both PRS and HRS (empty circles) to this baseline. The $p_{\rm T}$-average ratios are presented in Table \ref{tab:PRS_HRS_ratio}. It is observed that the PRS exhibits negligible impact on the $J/\psi$ yield across all production sources. In contrast, the HRS suppresses $J/\psi$ production from both the NRQCD and cluster collapse mechanisms, while leaving non-prompt $J/\psi$ yield largely unaffected. The total suppression effect of the HRS on the inclusive $J/\psi$ cross section is around 8$\%$. It is understandable according  to the following reasons. As shown in section \ref{sec:model}, in the HRS $J/\psi$ scatters with $p$, $n$, $\pi^{\pm}$, $\pi^0$, $\rho^{\pm}$, and $\rho^0$ inelastically, which will reduce the $J/\psi$ yields from the NRQCD and the cluster collapse. As the non-prompt $J/\psi$ is produced from the $b$-hadron decay, and we do not consider the $b$-hadron rescattering, the effect of HRS on the non-prompt $J/\psi$ yield is negligible. Moreover, in the PRS we do not include the scattering of $J/\psi$ with partons, such as 
$J/\psi+g\rightarrow c+\bar{c}$ and $J/\psi+\textrm{parton}\rightarrow c+\bar{c}+\textrm{parton}$ ($\textrm{parton}=g,q,\bar{q}$) \cite{Jpsi_parton_scat}. Thus the PRS
has little effect on the $J/\psi$ production.
%It is found that HRS has a reduction on the inclusive $J/\psi$ cross sections 

\section{\label{sec:conclusions}Conclusions}
We have investigated the inclusive $J/\psi$ production in pp collisions at $\sqrt{s} = 5.02$, 7, and 13 TeV using the PACIAE 4.0 model. This model, which extends PYTHIA 8.3 with partonic and hadronic rescatterings, incorporates the  NRQCD, cluster collapse, and non-prompt  production mechanisms. Our simulations reproduce the measured $p_{\rm T}$-differential cross sections at middle and forward rapidities. We find that the NRQCD channel provides the dominant contribution to the total yield, with sub-leading shares from the cluster collapse and non-prompt productions. As the collision energy rises, the relative contributions from the cluster‑collapse and non‑prompt processes increase, while the fraction from the NRQCD channel decreases. Furthermore, at forward rapidity the contributions from both the NRQCD and cluster‑collapse processes are enhanced relative to mid‑rapidity, whereas the non‑prompt component is suppressed. A decomposition of the NRQCD component reveals that the direct production together with the $\chi_{c2}$ feed‑down dominates the yield at low $p_{\rm T}$, while at high $p_{\rm T}$ the dominant contribution comes from the direct production combined with the $\chi_{c1}$ feed‑down. The feed‑down from $\psi(2S)$ and $\chi_{c0}$ accounts for about 5-8$\%$ and 2-3$\%$ of the NRQCD yield, respectively. We also observe clear energy and rapidity dependence in the fractional contributions of the different NRQCD sub‑channels. Furthermore, we find that the partonic rescattering has a negligible effect, while the hadronic rescattering selectively suppresses the  $J/\psi$ yields from the NRQCD and cluster collapse processes.

\begin{acknowledgments}

We would like to thank Prof. Qian Yang at Shandong University for his valuable discussions. This work is supported by the research fund from the School of Physics and Information Technology at Shaanxi Normal University, by the Scientific Research Foundation for the Returned Overseas Chinese Scholars, State Education Ministry, by Natural Science Basic Research Plan in Shaanxi Province of China (program No. 2023-JC-YB-012) and by the National Natural Science Foundation of China under Grant Nos. 11447024, 11505108 and 12375135.
\end{acknowledgments}

%\nocite{*}

%\bibliography{apssamp}% Produces the bibliography via BibTeX.

\end{document}